\documentclass[12pt]{iopart}
\pdfoutput=1
\usepackage{graphicx}

\usepackage{iopams}

\newcommand{\bra}[1]{\langle #1 |}
\newcommand{\ket}[1]{| #1 \rangle}

\newcommand{\anniop}[1]{\hat{a}_{#1}}

\begin{document}
\title{Photonic Qubits, Qutrits and Ququads accurately Prepared and Delivered on Demand}
\author{Peter B.\,R.\,Nisbet-Jones\footnote{Now at the National Physical Laboratory, Hampton Road, Teddington, Middlesex, UK, TW11 0LW}, 
Jerome Dilley, Annemarie Holleczek, Oliver Barter,
and Axel Kuhn}
\address{University of Oxford\\ Clarendon Laboratory\\ Parks Road, Oxford\\ OX1 3PU United Kingdom}
\ead{axel.kuhn@physics.ox.ac.uk}

\begin{abstract}
Reliable encoding of information in quantum systems is crucial to all approaches to quantum information processing or communication. This applies in particular to photons used in linear optics quantum computing (LOQC), which is scalable provided a deterministic single-photon emission and preparation is available. Here, we show that narrowband photons deterministically emitted from an atom-cavity system fulfill these requirements. Within their 500\,ns coherence time, we demonstrate a subdivision into {\bf\em d} time bins of various amplitudes and phases, which we use for encoding arbitrary qu-{\bf\em d}-its. The latter is done deterministically with a fidelity $> 95\%$ for qubits, verified using a newly developed time-resolved quantum-homodyne method.

\end{abstract}

\maketitle

\section{Introduction}
Using elementary quantum systems to encode information is the key to several novel approaches to quantum mechanical based computing \cite{DiVincenzo00, Ladd10}, with one successful technique being linear optics quantum computing (LOQC) in photonic circuits \cite{Knill01, Ralph01,Kok07, OBrien07,Pooley12}. Although this method is in principle scalable, in practice it is limited by the stochastic nature of spontaneous parametric down-conversion (SPDC) sources used to seed the circuits. These photon sources produce simultaneous pairs of photons, one is used as a herald, the other is used to initialize a qubit. Typically the photon is split by a $\eta:(1-\eta)$ beam splitter, leaving it in a superposition of two spatial modes, known as dual-rail encoding \cite{Pittman02:2}. This state can be described as
\begin{equation}
	  |\Psi\rangle=\alpha |10\rangle+\beta\,e^{i\phi}|01\rangle,
 \label{signal}
\end{equation}
where $\phi$ in the phase delay in the transmitted mode, $\ket{10}$ and $\ket{01}$ corresponding to the two spatial modes, and $\alpha=\sqrt{\eta}$ and $\beta = \sqrt{1-\eta}$. In the following, we extend this concept to the temporal modes \cite{Marcikic02} with $\ket{10}$ and $\ket{01}$ now representing a photon in the first or second time bin, respectively. Up until now, time-bin encoded photons have been obtained by ``pumping'' a non-linear crystal with a femtosecond pulse \cite{Marcikic02}, which had initially been passed through a unbalanced Mach-Zehnder interferometer. Eventually, time-bin entangled photons are created in, however, a completely probabilistic fashion.

Here, we demonstrate the deterministic initialisation and delivery of arbitrary time-bin encoded qubits, qutrits and even ququads in one single photon with the help of a strongly-coupled atom cavity system \cite{Kuhn10,Vasilev10,Nisbet11}. While qubits -- used as default information carriers in quantum information processing -- are encoded in the binary system, qutrits and ququads are their equivalent in the ternary and quaternary system, respectively. Without any increase in technological resources, we are able to encode information beyond the binary system with the help of our single photon source. Additionally, our photon generation process is inherently non-probabilistic, versatile, reconfigurable and is not subject to systematic photon losses. We verify the fidelity of the quantum state preparation with time-resolved quantum-homodyne measurements, performed by sending single signal and reference photons into an elementary photonic circuit. Photon correlations monitored in a time-resolved manner \cite{Legero06,Legero04} then allow the state to be partially reconstructed.

\section{Photon source and quantum-homodyne measurement} 

\begin{figure*}[t]
\begin{center}
\includegraphics[width=\textwidth]{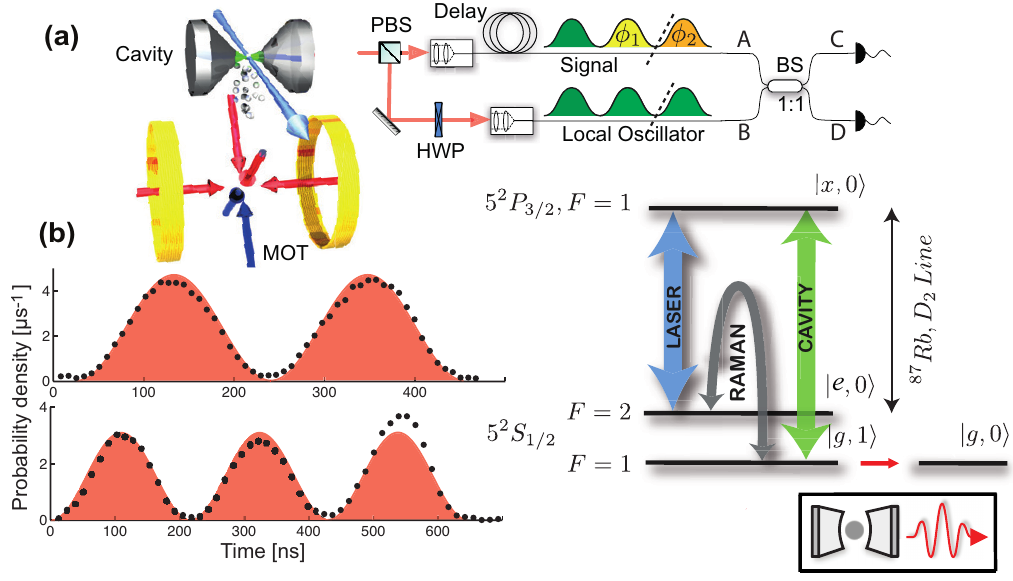}
\caption{Setup \textbf{(a)}: An atomic fountain injects $^{87}$Rb atoms into a high-finesse optical cavity, where Raman transitions between hyperfine ground states control the production of single photons of arbitrary spatio-temporal envelope. A $200\,\mathrm{m}$ fibre delays photons to arrive at a fiberized beam splitter together with subsequently emitted photons travelling the shorter path. The half-wave plate (HWP) is used to measure the quantum homodyne signal of parallely and perpendicularly polarized photons, respectively. \textbf{(b)} Probability density of twin- and triple-peak photons. The red-filled area indicates the desired $sin^2$-shaped probability density in every time bin.\label{fig:Scheme}}
\end{center}
\end{figure*}

We achieve the generation of photons with an a priori defined intensity envelope -- which was first demonstrated by Keller et al. \cite{Keller2004} -- with the help of a controlled Raman transitions on the D2-line of a strongly-coupled $^{87}$Rb atom-cavity system. Having refined the method \cite{Kuhn10,Vasilev10,Kuhn02} and experimentally verified that such photons remain identical \cite{Nisbet11}, our photons are produced with an efficiency of $\eta=85\,\%$  and a repetition rate of $1\,\mathrm{MHz}$, see Fig.\,\ref{fig:Scheme}a and Section \ref{methods}. In the following, we go one essential step further and demonstrate the complete control over the photons' phase profile.

 The amplitude of the driving pulse is modulated to produce $d$-peak photons of 230\,ns peak duration, which yields exactly one $\sin^2$-shaped intensity peak per time bin (or temporal mode). The probability density of twin-peak and triple-peak photons of 460\,ns and 690\,ns duration, respectively, are shown in Fig.\,\ref{fig:Scheme}b. The singleness of these photons is checked with the help of a  Hanbury-Brown-Twiss setup and the second order correlation function at zero detection-time difference is $g^2(\tau=0)<0.05$ \cite{Nisbet11}, which is the upper limit imposed by the shot noise of the detector dark counts. Our source also allows for imposing an arbitrary phase change of $\phi$ between neighbouring peaks within one photon. This is accomplished by applying a phase shift to the driving pulse using an acousto-optic modulator, which intrinsically maps $\phi$ onto the respective time bin. We emphasize that the Hamiltonian of the coupled system stays unitary throughout the whole process, and that any phase change is only imposed when the cavity mode is empty. This ensures an adiabatic evolution and prepares the emitted photon in a coherent superposition of temporal modes. The creation of complex states is now straight forward, as we can directly shape the amplitude and phase profile of the emitted single photons. 

To verify that the quantum state of the photon corresponds to the desired qubit, the relative phase between individual time bins has to be measured. This is achieved with a quantum-homodyne technique,  based on time-resolved two-photon interference measurements \cite{Legero06,Legero04}. In contrast to the continuous variable regime where a `classical' local oscillator (LO) \cite{Andersen2010} is used, the signal photon interferes on a beam splitter with a single LO photon of an intensity envelope identical to the signal photon, but no phase shift between time bins. For instance, a LO in state $|\Psi_{LO}\rangle = \left(\ket{10}+\ket{01}\right)/{\sqrt{2}}$ is used to analyse twin-peak photons encoding qubits. The photons arrive simultaneously at the beam splitter in spatio-temporal modes $A$ and $B$, such that the input state $\hat{a}_{A}^{\dagger}\hat{a}_{B}^{\dagger}|00\rangle$ translates into a superposition of output modes $C$ and $D$ using the operator relation of the beam splitter, $\hat{a}_{A,B}^{\dagger}\propto\hat{a}_{C}^{\dagger}\pm\hat{a}_{D}^{\dagger}$. Applied to twin-peak signal and LO photons, the unnormalised creator of the quantum state reads
\begin{eqnarray}
(\hat{a}_{C1}^{\dagger})^2-(\hat{a}_{D1}^{\dagger})^2 +e^{i\phi}\Big((\hat{a}_{C2}^{\dagger})^2-(\hat{a}_{D2}^{\dagger})^2\Big)+\label{Eqn:Out2}\\
(\hat{a}_{C1}^{\dagger}\hat{a}_{C2}^{\dagger}-\hat{a}_{D1}^{\dagger}\hat{a}_{D2}^{\dagger})(1+e^{i\phi})-(\hat{a}_{C2}^{\dagger}\hat{a}_{D1}^{\dagger}-\hat{a}_{D2}^{\dagger}\hat{a}_{C1}^{\dagger}) (1-e^{i\phi}),\nonumber
\end{eqnarray}
where the subscripts \{$1,\,2$\} refer to the time bin and \{$C,\,D$\} to the output port of the beam splitter. The first line contains the creation operator pairs where both photons are detected in the same time bin. In this case, the photons coalesce and are detected in the same output port.  The other terms describe the situation where the photons are detected in different time bins. Ideally, no cross-correlations are witnessed for identical photons $(\phi=0)$, which corresponds to the well-known Hong-Ou-Mandel effect \cite{Hong1987}. However, if we choose $\phi=\pi$,  the photons are  forced into opposite output ports whenever the detections occur in different time bins. The photon correlations, which normally follow bosonic statistics, then seem quasi-fermionic. This is explained by the first photo detection projecting the one remaining excitation into a mode superposition of well-defined mutual phase, which subsequently changes by $\pi$ until the second photon is detected. By examining the number of cross-correlations between different time bins, the relative phase difference, $\phi$, between any pair of time bins can be determined.

\begin{figure*}[t]
\begin{center}
\includegraphics[width=\textwidth]{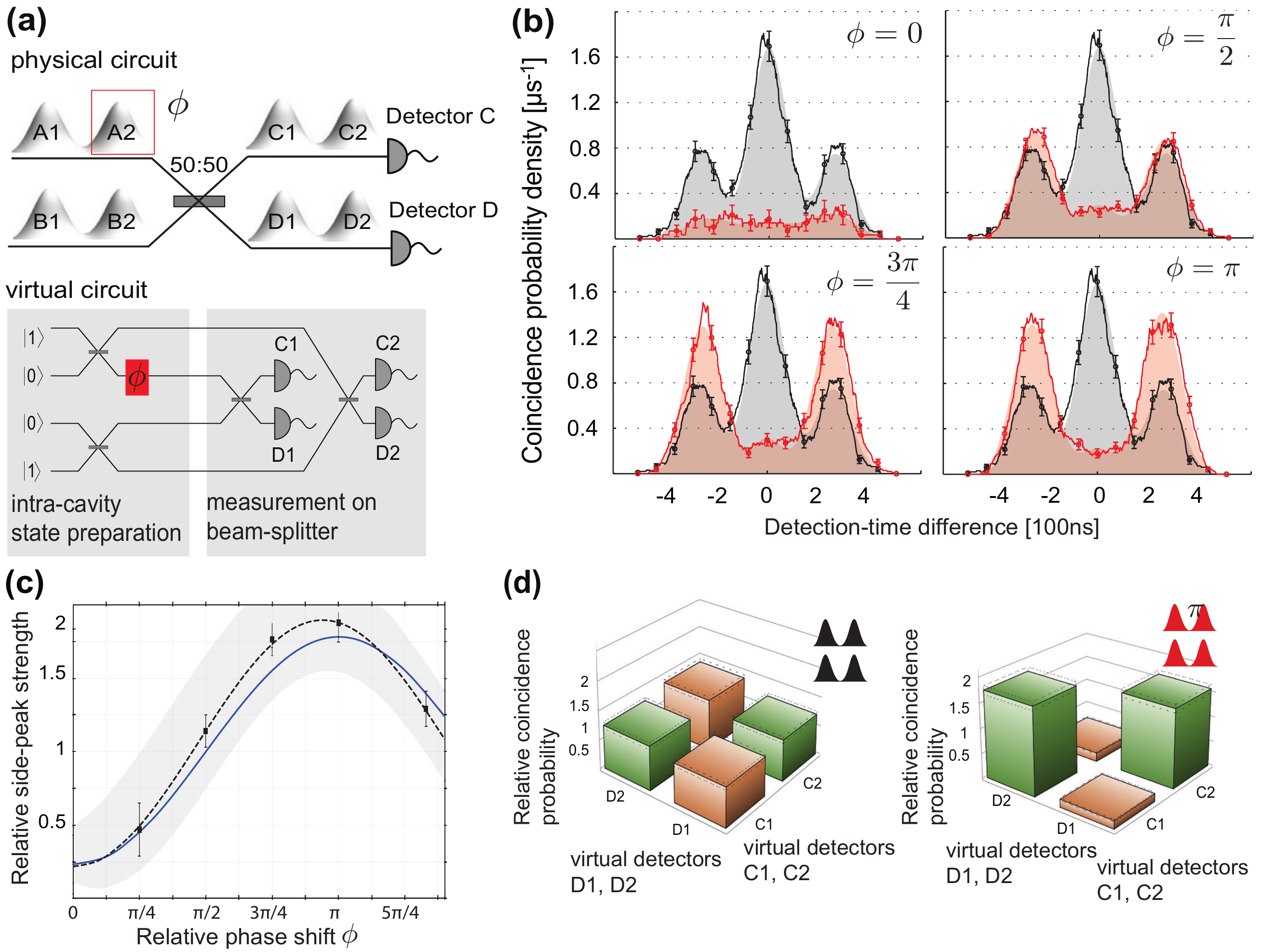}
\caption{Equivalent circuits: \textbf{(a)} Measurements on twin-peak photons arriving simultaneously at a beam splitter are equivalent to experiments in a virtual circuit with spatial modes corresponding to temporal modes. \textbf{(b)} Correlations from their homodyne measurement as a function of detection-time difference (grey: reference of perpendicular polarization; red: quantum-homodyne signal of parallel polarization, red- and grey-shaded areas: Best model from fitting three auto-correlated $sin^2$-pulses to the measured coincidence probabilities). The red and black solid traces result from summing all coincidences found within a 60\,ns wide interval around each point of the trace. For some of these densly spaced data points, the statistical error is shown. The phase between the signal photon peaks varies from 0 to $\pi$. \textbf{(c)} Relative strength of the cross-correlation side peaks, expectations (blue) and the $95\,\%$ confidence bounds (grey). The $0\rightarrow2$ range is not fully spanned due to coherence-time limitations, background counts and polarisation deficiencies. \textbf{(d)} Relative coincidence probabilities (RCP, see Section \ref{methods}) between virtual detectors:  The time resolution allows the mapping of real detector clicks to virtual detectors C1, C2, D1, and D2 firing. We use this to sub-divide the quantum-homodyne signal  into cases where both detectors fire in the same time bin (brown), or in successive time bins (green). The left coincidence matrix is for non-interfering perpendicularly polarized photons, the right matrix for photons of parallel polarisation with a $\pi$ phase shift between the time bins of the signal photon.\label{fig:Twin}}
\end{center}
\end{figure*}

\section{Experimental qu-d-it preparation and delivery}
\begin{figure*}[!t]
\begin{center}
\includegraphics[width=\textwidth]{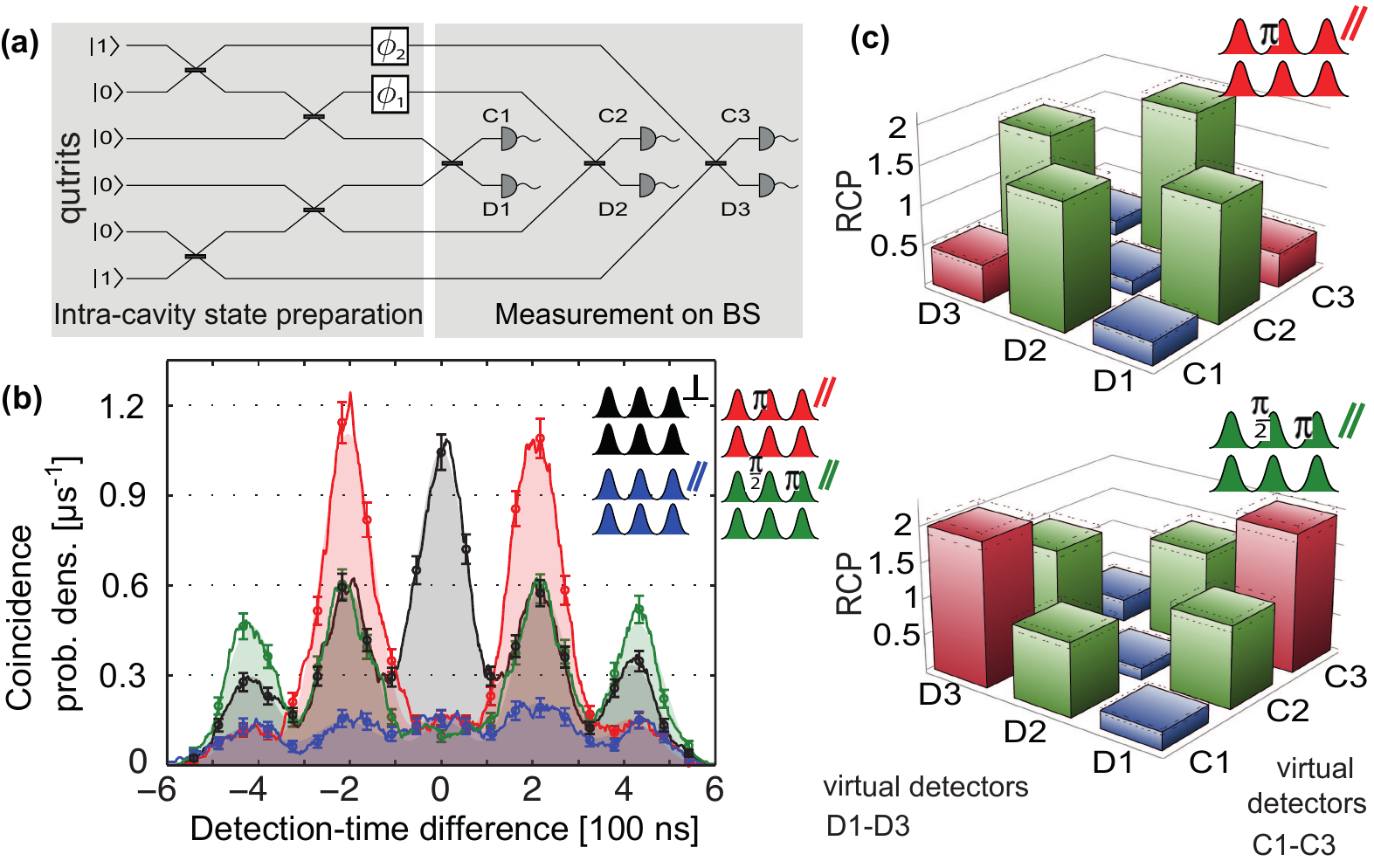}
\caption{Qutrits: \textbf{(a)} For triple-peak photons, the  virtual circuit shows 7 beam splitters and 6 detectors, whilst the physical apparatus is unchanged.  \textbf{(b)} Time-resolved homodyne signal  for photons of perpendicular (black) and parallel (blue) polarization, with the signal photon having a phase shift in the central time bin of $\phi_1=\pi$ (red) and with phase shifts of  $\phi_1=\pi/2$ in the central and $\phi_2=\pi$ in the final time bins (green). As before, the solid traces result from summing all coincidences found within a 60\,ns wide interval around each point of the trace. For some of these data points, the statistical error is shown. The shaded areas under the plotted lines correspond to the fit of five auto-correlated $sin^2$-pulses to the measured coincidence probabilities.
\textbf{(c)} Relative coincidence probabilities (RCP, see Section \ref{methods}) between virtual detectors (blue: detections within the same time bin; green: detections in successive time bins; red: detections two time-bins apart). \label{fig:Trip}}
\end{center}
\end{figure*}

\begin{figure*}[!t]
\begin{center}
\includegraphics[width=\textwidth]{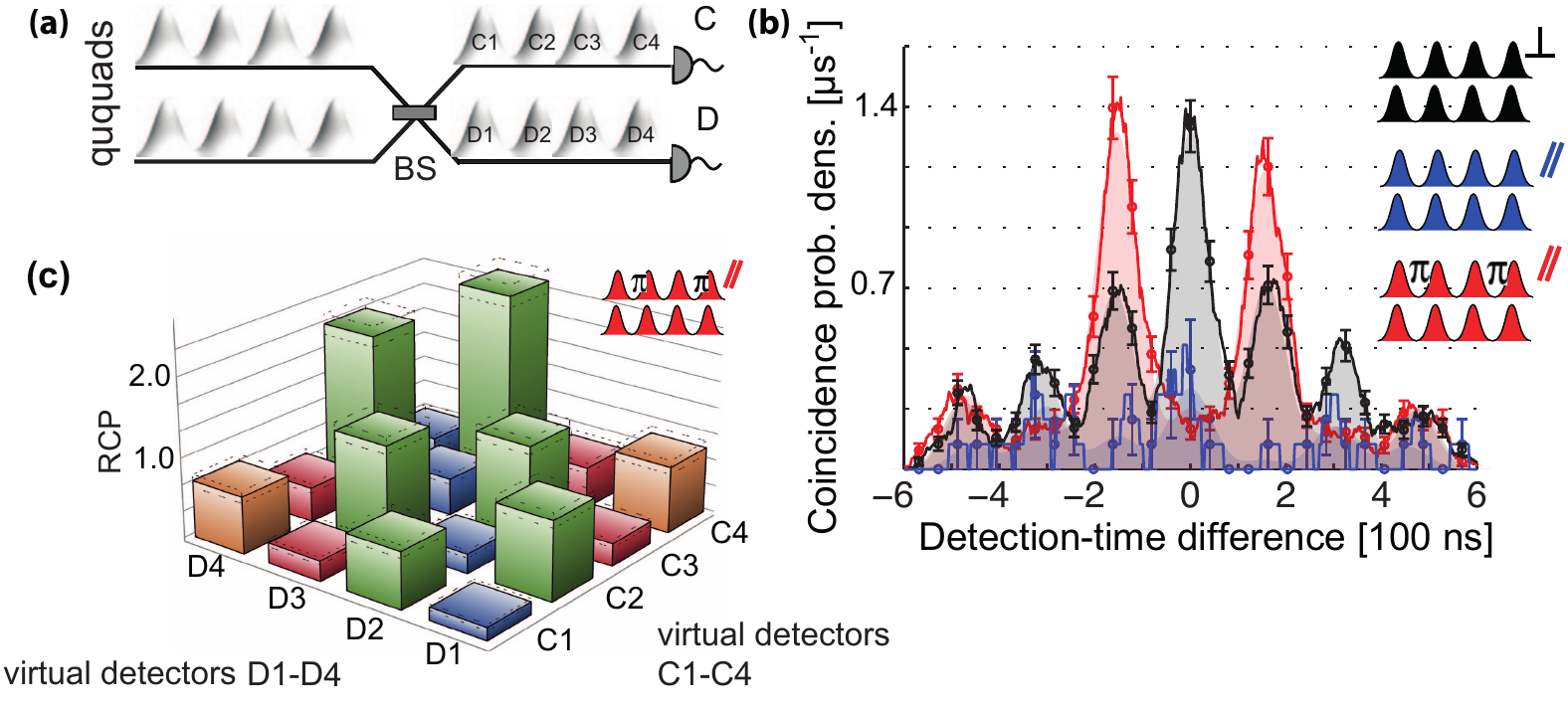}
\caption{Ququads: \textbf{(a)} encoded in quad-peak photons and their \textbf{(b)} resolved homodyne signal of perpendicular (black), parallel (blue) polarization, and for phase shifts of $\phi_1=\phi_3=\pi$ in the signal photon (red). As before, the solid traces are obtained from summing all coincidences found within an 60\,ns wide interval around each point of the trace and for some of these data points, we show the statistical error. The shaded areas in the plot correspond to the fit of seven auto-correlated $sin^2$-pulses to the measured coincidence probabilities. \textbf{(c)} RCP between the eight virtual detectors in this latter case.\label{fig:Quad}}
\end{center}
\end{figure*}

An outline of the  arrangement is shown in Fig.\,\ref{fig:Scheme}a and the photon probability density for both the qubit and qutrit in Fig.\,\ref{fig:Scheme}b.  The stream of alternating signal and LO photons is emitted from the cavity and split by a polarizing beam splitter (PBS) between long (upper) and short (lower) beam paths, causing successively emitted photons to arrive simultaneously at the non-polarizing 50:50 beam splitter (see Section \ref{methods}). Perpendicularly polarized photons do not interfere, but as their envelopes are identical, their cross-correlation function reflects the auto-correlation of the photon's shape. For twin-peak photons, a triple-peak convolution is witnessed (grey in Fig.\,\ref{fig:Twin}b), with a large central peak caused by detector clicks in identical time bins which are represented by the annihilation operators $\anniop{C1}\anniop{D1}$ and $\anniop{C2}\anniop{D2}$. The two satellite peaks of equal height arise from cross correlations between time bins represented by $\anniop{C1}\anniop{D2}$ and $\anniop{C2}\anniop{D1}$, respectively. 
With parallel polarization, the photons do interfere and the cross-correlation function (red) varies with the relative phase between the peaks of the signal photon. As discussed above, no correlations are found within identical time bins so the central peak vanishes. This is not the case for the satellite peaks at $\pm 230\,$ns, where the number of counts depends strongly on $\phi$ and is expected to vary between zero and twice the reference value found with perpendicularly polarized photons. We also note that a residual signal is always found in the coincidence measurements of parallely polarized photons (Fig. \ref{fig:Twin}b for qubits, Fig. \ref{fig:Trip}b for qutrits and Fig. \ref{fig:Quad}b for ququads, respectively). From photon statistics and the dark count rate of the detectors, half the background can be attributed to correlations involving detector dark counts while the remainder can be explained by a polarization or mode mismatch of the interfering photons due to a residual birefringence of the optical fibres. Dephasing effects play a minor role, as the residual signal does not vary significantly with time \cite{Legero06}. 

Fig.\,\ref{fig:Twin}c shows the strength of these satellite peaks relative to the reference signal as a function of $\phi$; the expected sinusoidal behaviour is exhibited, with minima and maxima witnessed at $\phi=0$ and $\phi=\pi$, respectively. Taking the 500\,ns coherence time into account, the expected change in peak strength has been calculated (solid trace) and is found to agree well also with the data (dashed) within the $95\,\%$ confidence bounds of a $\chi^2$-test. The limitation of the coherence time of the photons to 500\,ns is mainly due to the used laser system and the drift in the cavity resonance frequency, as the cavity is not actively locked while the atom cloud is flying through its centre. Furthermore, changes in the magnetic environment beyond our control also reduce the coherence time.

Greater insight can be obtained from the equivalent `virtual' photonic circuit in Fig.\,\ref{fig:Twin}a where the various time bins (or temporal modes) are represented as if they were spatial modes. Two photons are generated, each existing in a superposition of two time bins with a possible phase shift,  interfering on a single beam splitter with detectors $C$ and $D$ in the output ports. This beam splitter is used twice -- once per time bin -- and so the equivalent virtual circuit consists of  four beam splitters, where two mimic the state preparation. Due to the photons being much longer than the detectors' time resolution, we are able to distinguish whether individual photons are detected in the first or second time bin, respectively. We therefore associate the two physical detectors $C$ and $D$ with four virtual detectors, $C1, C2, D1,$ and $D2$, where for example the virtual detector $C1$ stands for the physical detector $C$ measuring events only in the first time bin. We then retrieve the four cross-correlation signals between all virtual detectors,  $\anniop{C1}\anniop{D1}$ and $\anniop{C2}\anniop{D2}$ (same time bin), or $\anniop{C1}\anniop{D2}$ and $\anniop{C2}\anniop{D1}$ (different time bins) using the absolute time of each photodetection along with the information \emph{which} virtual detector fired. 
We also use the virtual-detector concept for detecting qutrits (six virtual detectors) and ququads (eight virtual detectors). The respective data can be found in Fig.\,\ref{fig:Trip}c and Fig.\,\ref{fig:Quad}c.
Fig.\,\ref{fig:Twin}d shows these correlation matrices for twin-peak photons of perpendicular and parallel polarization,  with a $\phi=\pi$ phase shift in the signal photon. By comparing the correlations to  expectations, a qubit-preparation fidelity of $F=0.96\pm 0.01$ is found, which is deduced in Section \ref{methods}.

The control of the photon-generation process goes far beyond simple qubits in time-bin encoding. We successfully added more time bins of arbitrary amplitudes and relative phases, thus producing a large variety of arbitrary qudits. 
In Fig.\,\ref{fig:Trip} and \ref{fig:Quad} the correlations of triple- and quad-peak photons, representing the states
\begin{eqnarray}
\ket{\Psi_3} \propto \ket{100}+e^{i\phi_1}\ket{010}+e^{i\phi_2}\ket{001},\nonumber\\
\ket{\Psi_4} \propto\ket{1000}+e^{i\phi_1}\ket{0100}+e^{i\phi_2}\ket{0010}+e^{i\phi_3}\ket{0001},
\end{eqnarray}
are shown. The virtual circuit in Fig.\,\ref{fig:Trip}a illustrates the preparation and analysis of a qutrit, $\ket{\Psi_3}$. Although the physical apparatus is unchanged, it now mimics six detectors C1-3 and D1-3. Fig.\,\ref{fig:Trip}b and \ref{fig:Quad}b show the time-resolved correlation signal for qutrits and ququads encoded in triple- and quad-peak photons, respectively, which we also present as correlations between virtual detectors in Fig.\,\ref{fig:Trip}c and \ref{fig:Quad}c. For the 920\,ns long quad-peak photons, the probabilities for $\anniop{C1}\anniop{D4}$ and $\anniop{C4}\anniop{D1}$ correlations are small and others fluctuate due to exceeding the 500\,ns coherence time. Also the state-preparation fidelity drops from 0.94 for qutrits to 0.89 for ququads (see Section 5). Hence only within the coherence time,  the probabilities for observing correlations between time bins depend on the relative phases as predicted. Further increasing the state-vector dimension seems nonetheless feasible. About eight well-separable time bins could be produced within the coherence span, as the dynamics of the atom-cavity system is only restricted by the atom-cavity coupling strength,  $g_0 = 2\pi\cdot 15\,$MHz \cite{Nisbet11}. 

\section{Discussion and Outlook}

The novel approach we have implemented allows for the preparation and deterministic delivery of single photons in arbitrary quantum superpositions of various temporal modes. We made use of this sub-division into time bins to reliably encode a selection of qudits in single photons. Photon-generation efficiencies up to $\eta=85\,\%$ and state-preparation fidelities of $F=96\%$ have been found by quantum homodyning (see Section \ref{methods}). Time-bin encoding takes place upon photon generation which is inherently non-probabilistic, versatile, reconfigurable and not subject to systematic photon losses. These are substantial advantages if compared to other approaches that are probabilistic if based on SPDC and also subject to intrinsic photon loss if they rely on variable amplitude attenuation \cite{Kolchin08}, phase modulation \cite{Specht09} or interferometric techniques for photon shaping \cite{Marcikic02}. The availability of time bins as an additional degree of freedom to LOQC in an essentially deterministic photon-generation scheme is a big step towards large-scale quantum computing in photonic networks \cite{Bennett08}. 
Arbitrary single-qubit operations on time-bin encoded qubits seem straightforward to implement with phase-coherent optical delay lines and active optical routing to either switch between temporal and spatial modes, or to swap the two time bins. 
Moreover, the probability of having $N$ photons simultaneously available, either from many atom-cavity sources or transmitted through a photon-delay network, scales like $\eta^N$, with $\eta \gg g^{(2)}(0)$. Restrictions as for non-heralded SPDC sources (with $\eta\simeq g^{(2)}(0)$) do  not apply.   Controlling the atom-photon coupling might also allow the mapping of atomic superposition states to time-binned photons \cite{Dilley12,Ritter12}; and the long coherence time, combined with fast detectors, makes  real-time feedback possible during photon generation.

\section{Methods}
\label{methods}

\textbf{Photon source and efficiency:} The method we use for generating and shaping individual photons is theoretically discussed and introduced in \cite{Vasilev10}, and a detailed description of our photon source is to be found in \cite{Nisbet11}. It is based on a single $^{87}$Rb atom coupled to a high-finesse cavity undergoing a Raman transition between two of its hyperfine ground states, labeled $|e\rangle$ and $|g\rangle$ in Fig.\,1a, via an excited state $|x\rangle$. This transition -- driven by a laser pulse with Rabi frequency $\Omega(t)$, and the vacuum field of the cavity mode with coupling strength $g_0$  --  results in the emission of a single photon through the output coupling mirror of the cavity. Atoms are injected into the cavity using an atomic fountain, and couple strongly to the cavity mode for a maximum duration of 200\,$\mu$s. This is long enough to trigger up to 200 subsequent photon emissions. For the most strongly coupled atoms, photons are detected for $20\%$ of the driving pulses. Taking the detector efficiencies, fibre-coupling losses and cavity losses into account, this corresponds to an intra-cavity photon generation efficiency of $85\%$, of which $60\%$ leave our cavity through its output coupling mirror. Increasing the latter to $90\%$ is straightforward if one can afford better mirrors \cite{Ritter12}. 

\textbf{Quantum homodyne setup:} The photon source produces an alternating sequence of unpolarized LO and signal photons. A polarizing beam splitter randomly directs these into either a $1\,\mu$s-long delay line (200\,m optical fibre), or directly to the final beam splitter. The fibre-imposed delay matches the time distance between photon productions, such that any pair of successively generated photons arrives simultaneously at the beam splitter with a probability of $25\,\%$. If polarized photons were used in combination with fast switching optical elements, this could be increased to $100\,\%$.
All photo-detection events in the output ports of the beam splitter get recorded with sub-ns time resolution. We use the recorded data stream to determine photon-photon correlations either as a function of the detection-time difference, or within (or between) selected time bins. This allows for monitoring the continuous phase evolution within a photon and for the detection of discrete phase jumps between the time bins of interest. For simultaneously arriving photon pairs, 
\textbf{the coincidence probability density} $\times dt$ gives the probability of detecting a coincidence count in an interval of width $dt$ around the specified detection-time difference. Any trace shown here is normalized to the corresponding mean number of photon pairs that we retrieve during the very same experimental run from the amount of coincidences between non-interfering photons of detection-time delay $>1\,\mu$s (the periodicity of the source).    

\textbf{The relative coincidence probability (RCP)} is defined by the ratio of quantum-homodyne correlations to the number of reference correlations with perpendicularly polarized and therefore non-interfering photons. In the latter case, the photons randomly split. This implies that the RCP ranges at most from zero (photon coalescence, no correlations) to two (photon anti-coalescence, correlations in any case).

\textbf{Fidelity of the qubit preparation:}  To determine the fidelity of the state preparation, the density matrix $\hat{\sigma}$ has to be reconstructed. For a qubit of $\phi=\pi$, one gets
\begin{equation}
\hat{\sigma} = \frac{1}{2} \left(\begin{array}{ccc}1.01 & -0.85 \\ -0.85 & 0.99\end{array}\right).
\end{equation}
The values on the diagonal are obtained using non-interfering photons of perpendicular polarization in the two-photon coincidence measurement (Fig. 2d). The numbers are proportional to the square root of the number of correlations detected in time bin $1$ and $2$, respectively, and then normalised to one. We further assume that signal and LO photons only differ in phase, such that their off-diagonal elements only differ in sign. Their magnitude is given by the ratio of the square root of the maximum observed to the maximum possible side peak visibility.
Having obtained the density matrix of the single photon state, the fidelity \cite{NielsenChuang} itself is is calculated with $F(\ket{\psi},\hat{\sigma})=\sqrt{\bra{\psi} \hat{\sigma} \ket{\psi}}=0.96\pm 0.01$, where $\ket{\psi} = (\ket{10}-\ket{01})/\sqrt{2}$ is the reference state,  $\hat{\sigma}$ is the reconstructed density matrix, and the error results from propagating the standard deviation associated with the absolute number of $N=418$ correlations forming these side peaks. An equivalent procedure yields fidelities of $0.94\pm 0.01$ and $0.89\pm 0.02$ for qutrits and ququads, respectively.

\section*{Acknowledgements}
This work was supported by the Engineering and Physical Sciences Research Council (EP/E023568/1 and QIPIRC) and the Deutsche Forschungsgemeinschaft (DFG, RU 635).

\section*{References}
\bibliographystyle{unsrt.bst}

\begin{thebibliography}{10}

\bibitem{DiVincenzo00}
D. P. DiVincenzo, 
\newblock{The Physical Implementation of Quantum Computation}.
\newblock{\em Fortschr. Phys.} $\mathbf{48,}$ 771 (2000).

\bibitem{Ladd10}
T. D. Ladd et al. 
\newblock{Quantum computers}. 
\newblock{\em Nature} $\mathbf{464,}$ 45--53 (2010).

\bibitem{Knill01}
E. Knill, R. Laflamme, G. J. Milburn,
\newblock{A scheme for efficient quantum computing with linear optics}. 
\newblock{\em Nature} $\mathbf{409,}$ 46--52 (2001).

\bibitem{Ralph01}
T.~C. Ralph, A. G. White, W. J. Munro, G. J. Milburn, 
\newblock{Simple scheme for efficient linear optics quantum gates}. 
\newblock{\em Phys. Rev. A} $\mathbf{65,}$ 012314 (2001).

\bibitem{Kok07}
P. Kok, K. Nemoto, T. C. Ralph, J. P. Dowling, G. J. Milburn, 
\newblock{Linear optical quantum computing with photonic qubits}. 
\newblock{\em Rev. Mod. Phys.}  $\mathbf{79,}$ 135--174 (2007).

\bibitem{OBrien07}
J.~L. O'Brien, 
\newblock{Optical Quantum Computing}. 
\newblock{\em Science} $\mathbf{318,}$ 1567--1570 (2007).


\bibitem{Pooley12}
M. A. Pooley, et al.
\newblock{Controlled-NOT gate operating with single photons}. 
\newblock{\em Appl. Phys. Lett.} $\mathbf{100,}$ 211103 (2012).



\bibitem{Pittman02:2}
T.~B. Pittman, B.~C. Jacobs, J.~D. Franson, 
\newblock{Demonstration of Non-Deterministic Quantum Logic Operations using Linear Optical Elements}. 
\newblock{\em Phys. Rev. Lett.} $\mathbf{88,}$ 257902 (2002).

    
\bibitem{Marcikic02}
I. Marcikic et al. 
\newblock{Time-bin entangled qubits for quantum communication created by femtosecond pulses}. 
\newblock{\em Phys. Rev. A} $\mathbf{66,}$ 062308 (2002).

\bibitem{Kuhn10}
A. Kuhn, D. Ljunggren, 
\newblock{Cavity-based single-photon sources}. 
\newblock{\em  Contemp. Phys.} $\mathbf{51,}$ 289--313 (2010).

\bibitem{Vasilev10}
G.~S. Vasilev, D. Ljunggren, A. Kuhn, 
\newblock{Single photons made-to-measure}. 
\newblock{\em New J. Phys.} $\mathbf{12,}$ 063024 (2010).

\bibitem{Nisbet11}
P.~B.~R. Nisbet-Jones, J. Dilley, D. Ljunggren, A. Kuhn, 
\newblock{Highly efficient source for indistinguishable single photons of controlled shape}. 
\newblock{\em New J. Phys.} $\mathbf{13,}$ 103036 (2011).

\bibitem{Legero06}
T. Legero, T. Wilk, A. Kuhn, G. Rempe, 
\newblock{Characterization of single photons using two-photon interference}. 
\newblock{\em Adv. At. Mol. Opt. Phys.} $\mathbf{53,}$ 253--289 (2006).

\bibitem{Legero04}
T. Legero, T. Wilk, M. Hennrich, G. Rempe, A. Kuhn,
\newblock{Quantum beat of two single photons}. 
\newblock{\em Phys. Rev. Lett.} $\mathbf{93,}$ 070503 (2004).

\bibitem{Keller2004}
M. Keller, B. Lange, K. Hayasaka, W. Lange, H. Walther, 
\newblock{Continuous generation of single photons with controlled waveform in an ion-trap cavity system}. 
\newblock{\em Nature} $\mathbf{431,}$ 337--354 (2004).

\bibitem{Kuhn02}
A. Kuhn, M. Hennrich, G. Rempe, 
\newblock{Deterministic single-photon source for distributed quantum networking}. 
\newblock{\em Phys. Rev. Lett.} $\mathbf{89,}$ 067901 (2002).
  


\bibitem{Andersen2010}
U.~L. Andersen, G. Leuchs, C. Silberhorn, 
\newblock{Continuous Variable Quantum Information Processing}. 
\newblock{\em Laser \& Photonics Reviews} $\mathbf{4,}$ 1075--1078 (2010).

\bibitem{Hong1987}
C. K. Hong, Z. Y. Ou, L. Mandel,
\newblock{Measurement of subpicosecond time intervals between two photons by interference}. 
\newblock{Phys. Rev. Lett.} $\mathbf{59,}$ 2044--2046 (1987).

\bibitem{Kolchin08}
P. Kolchin,  C. Belthangady, S. Du, G. Y. Yin, and S. E. Harris
\newblock{Electro-Optic Modulation of Single Photons} 
\newblock{\em Phys. Rev. Lett.} $\mathbf{101,}$ 103601 (2008).

\bibitem{Specht09}
H. P. Specht et al. 
\newblock{Phase shaping of single-photon wave packets}. 
\newblock{\em Nat. Photonics} $\mathbf{3,}$ 469--472 (2009).


\bibitem{Bennett08}
A. J. Bennett et al.
\newblock{Experimental position-time entanglement with degenerate single photons}   
\newblock{\em Phys. Rev. A} $\mathbf{77,}$ 023803 (2008).

\bibitem{Dilley12}
J. Dilley, P. Nisbet-Jones, B. W. Shore, A. Kuhn, 
\newblock{Single-photon absorption in coupled atom-cavity systems}. 
\newblock{\em Phys. Rev. A} $\mathbf{85,}$ 023834 (2012).

\bibitem{Ritter12}
S. Ritter et al. 
\newblock{An elementary quantum network of single atoms in optical cavities}. 
\newblock{\em Nature} $\mathbf{484,}$ 195 (2012).
   
\bibitem{NielsenChuang}
M. A. Nielsen, I. L. Chuang, 
\newblock{\em Quantum Computation and Quantum Information} (Cambridge University Press, 2000).


\end{thebibliography}

\end{document}